\documentclass[11pt,twoside]{article}

\usepackage{asp2006}
\usepackage{epsf}
\usepackage{psfig}
\usepackage{lscape}

\begin{document}
\title{X-ray observations of NGC 1365: time-resolved eclipse of the X-ray source}   
\author{G. Risaliti}   
\affil{INAF-Arcetri Observatory, Largo E. Fermi 5, Firenze, ITALY, \& Harvard-Smithsonian Center for Astrophysics, 60 Garden Street, Cambridge MA, USA}    

\begin{abstract} 
We present an extraordinary variation of the X-ray spectrum of the obscured AGN in NGC 1365, which was 
observed by Chandra to change from Compton-thin to Compton-thick and back to Compton-thin in four days.
This fast variation imply a a size of $\sim10^{14}$~cm for the emitting region,
 and an extremely compact size ($\sim10^{16}$~cm)
of the clumpy circumnuclear absorber.
\end{abstract}


\section{Introduction}   
NGC 1365 is a nearby (D=19 Mpc) obscured Seyfert Galaxy, with an
interesting observational history in the X-rays: it has been observed 
with ASCA (Iyomoto et al.~1997), BeppoSAX (Risaliti et al.~2000),
Chandra (Risaliti et al.~2005A) and 5 times with XMM-Newton (Risaliti et al.~2005A,
2005B). These observations revealed dramatic changes in its spectral state, from
reflection-dominated (characterized by a flat, faint continuum and a prominent
iron emission line at 6.4 keV, with equivalent width EW$>$1~keV) to transmission
dominated, characterized by a strong power law continuum absorbed by a column density
$N_H=1.5-5\times10^{23}$~cm$^{-2}$. The fastest variations occurred within 
as short a time as three weeks, between the Chandra observation (reflection dominated)
and the first XMM-Newton observation (transmission-dominated), and again between the first
and the second XMM-Newton observations.
In principle, such extreme variations can be due to two physical phenomena:
absorption variability, due to a Compton-thick cloud, with $N_H>10^{24}$~cm$^{-2}$,
obscuring the X-ray source, or a switch off of the primary X-ray emission (in this
case the reflection component would be visible for some time, due to the
longer path to the observer of the reflected light rays).
 
The short observed timescales suggest that the correct scenario is the former,
since a complete switch off of the X-ray source in times shorter than
three weeks is unlikely (see discussion in Risaliti et al.~2005A).
Moreover, it suggests that the obscuring cloud is located at a distance D$<0.1$pc
from the center, much smaller than the standard parsec-scale torus assumed
in Unifed Models (e.g. Krolik \& Begelman 1988).

In order to investigate the Compton-thick/Compotn-thin changes at shorter timescales,
we monitored NGC~1365 with six short (15~ks) observations performed once every two days
from April 14 to April 24, 2006.
Here we report the results of this monitoring, which showed a spectral change
from transmission to reflection dominated in two days, and discuss the implication
of this extreme variability for the estimate of the size of the X-ray source and
the location of the obscuring gas.
 
\section{Chandra observation of the X-ray eclipse}

The Chandra observations were reduced and analyzed with a standard
procedure using the CIAO 3.3 package. The spectra of the first three 
observations are shown in Fig.~1, while the most relevant parameters of
the analysis of the six spectra are summarized in Table~1.
The main result of our analysis is apparent from Fig.~1: the second observation,
performed two days after the first one, and two days before the third one,
caught the source in a clearly reflection-dominated state. The spectral fitting
confirms the visual analysis: all the spectra except for the second one are well fitted
by a soft themal emission with temperature $kT\sim0.8$~keV (constant within 2\% in all 
observations), an absorbed power law with a constant photon index, and an iron emission
line, also compatible with being constant in all observations, with equivalent width 
$EW\sim150$~eV.
The second spectrum is instead well fitted by the same soft component as the
other spectra, plus a cold reflection continuum and an iron line with the same
flux as in the other spectra, and therefore with a much higher equivalent
width ($EW>1$~keV), due to the lower continuum level. 

\begin{figure}[h]
\psfig{figure=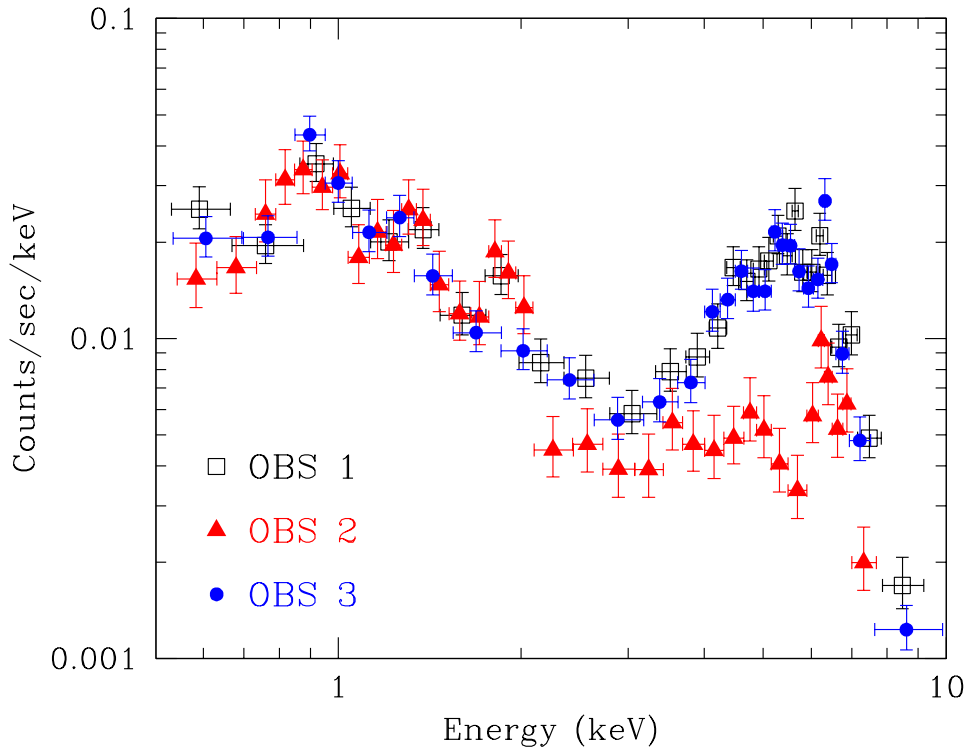,width=13cm}
\caption{Spectra from the first three Chandra observations of NGC 1365, 
taken once every two days.
The second observation caught the source in a reflection-dominated state,
while the other spectra show the intrinsic emission absorbed by a column density
$N_H\sim\times10^{23}$~cm$^{-2}$. The following three observations provided spectra
similar to the third one, and are omitted for clarity.}
\end{figure} 

\begin{table}
\centerline{\begin{tabular}{lcccc}
\hline
OBS & $\Gamma$ & $N_H^a$ & $N_{H,2}^b$ & $N^c$ \\
\hline
OBS 1 & 1.4$^{+1.1}_{-0.4}$ & 46$^{+11}_{-12}$ & 40$^{+6}_{-3}$ & 3.7$^{+0.8}_{-0.7}$ \\
OBS 2 & --                  & $>100$           & $>100$       & --                    \\
OBS 3 & 2.7$^{+1.5}_{-1.0}$ & 60$^{+19}_{-13}$ & 49$^{+6}_{-6}$ & 3.1$^{+0.7}_{-0.6}$ \\
OBS 4 & 1.8$^{+0.7}_{-0.6}$ & 36$^{+7}_{-7}$   & 34$^{+3}_{-3}$ & 4.3$^{+0.5}_{-0.5}$ \\
OBS 5 & 2.0$^{+0.5}_{-0.6}$ & 44$^{+7}_{-7}$   & 41$^{+3}_{-3}$ & 5.6$^{+0.7}_{-0.6}$ \\
OBS 6 & 1.4$^{+0.4}_{-0.2}$ & 22$^{+3}_{-2}$   & 23$^{+1}_{-1}$ & 4.6$^{+0.2}_{-0.3}$ \\
\hline
\end{tabular}}

\caption{\footnotesize{NGC 1365 - Spectral fittings.
$^a$: Column density in units of 10$^{22}$~cm$^{-2}$, obtained fitting a model with a
free photon index power law.
$^b$: Same, with the photon index frozen to the average value, $<\Gamma>=1.64$.
$^c$: Normalization of the power law, in units of 10$^{-3}$ keV~s$^{-1}$~cm$^{-2}$~keV$^{-1}$.}}
\end{table}

We interpret the observed variation as due to a Compton-thick cloud moving with Keplerian velocity
across the line of sight of the X-ray source. A schematic description of the occultaion event is
shown in Fig.~2.

Using two days as upper limit for the occultation time, we can estimate the dimension of the emitting
source. A first estimate can be obtained assuming a velocity of the obscuring cloud.
We note that the source has been caught in a Compton-thick state in $\sim$1/3 of the past observations.
If the obscuring clouds cover $\sim$1/3 of the line of sight, they are expected to be an efficient
mirror for the intrinsic X-ray emission. Therefore, we expect that a significant fraction -if not all-
of the observed reflected component (continuum plus narrow iron line) originates in these clouds.
Therefore, we can use the upper limit to the width of the iron line in the Compton thick spectra as an
estimate of the order of magnitude of the cloud velocity. From the XMM 2 observation (the one with the
highest S/N among Compton-thick spectra) we obtain $V<7,000$~km~s$^{-1}$.
Assuming a cloud velocity $V_K=7\times10^8$~cm~s$^{-1}$ we obtain that the dimension of the X-ray source
is $D_S\sim10^{14}$~cm.
\begin{figure}[h]
\psfig{figure=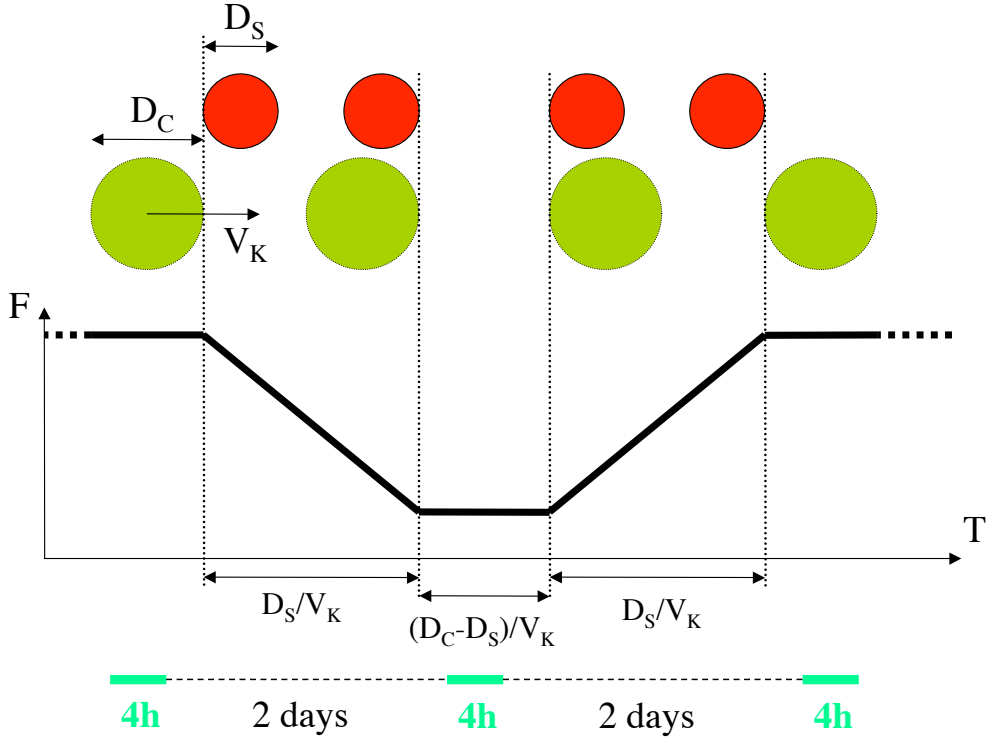,angle=-90,width=13cm}
\caption{\footnotesize{
Schematic representation of the observed eclipse. The intervening thick cloud 
(green circle, diamater $D_C$) vith Keplerian velocity $V_K$ starts covering 
the X-ray source (red circle, diameter $D_S$) at some time between the first 
and second Chandra observations. After a time  $T_1= D_S/V_K$ the source is 
completely covered, and remains obscured for a time $T_2=(D_C-D_S)/V_K$. 
In this state it is observed for the second time by Chandra. Then, it gradually 
uncovers, until it is back in the initial state. After some more time it is 
observed again by Chandra. From the times between the observations we infer 
$T_2+2T_1 < 4$ days; $T_2 > 4$ hours. The lower part of the figure shows the 
time evolution of the observed flux.
}}
\end{figure} 

The available mass estimates for the black hole in NGC~1365 are $M_{BH}=4_{-2}^{+4}\times10^7~M_\odot$ from
the $\sigma$-BH mass relation of Tremaine et al.~2002, using the dispersion estimated by Oliva et al.~1995
and $M_{BH}=7_{-3}^{+6}\times10^7~M_\odot$ from the relation between BH mass and K luminosity of the bulge
(Marconi \& Hunt 2003), with K luminosity measured in the 2MASS survey (Dong \& de Robertis 2006).
Based on these values, the above estimate of $D_S$ corresponds to 3-20 gravitational radii $R_G$. 
Assuming the average value, $D_S\sim10~R_G$.
In order to have a complete obscuration of the source, the cloud size $D_C$ must be the same or
larger than
the X-ray emitting source. Assuming a constant cloud density this implies a density
 $\rho_C\sim10^{10}N_{H,24}$~cm$^{-3}$, where $N_{H,24}$ is the column density in units of 10$^{24}$~cm$^{-2}$.
This is the density expected for a broad line cloud, consistently with the adopted velocity.
We conclude that a consistent scheme can be obtained if the source occultation is due to a broad line
region cloud, and the X-ray emitting region has a linear dimension of 10$^{14}$~cm.

\acknowledgements 
We acknowledge financial contribution from contract ASI-INAF I/023/05/0 and from
NASA grant GO6-7102X. 

\end{document}